\documentclass{article}

\usepackage{graphicx}
\usepackage{amsmath}

\usepackage{algorithm,amsfonts,amssymb,amsthm}
\usepackage{mathrsfs}
\usepackage{appendix}
\usepackage{url}

\usepackage{cleveref} 
\creflabelformat{equation}{#2(#1)#3}
\crefname{equation}{}{}

\crefname{figure}{Fig.}{}
\crefname{table}{Table}{}
\crefname{section}{Section}{}

\usepackage{ascmac}

\usepackage{mathrsfs}
\usepackage{mathtools}

\makeatletter
\newcommand{\subsubsubsection}{\@startsection{paragraph}{4}{\z@}%
  {1.0\Cvs \@plus.5\Cdp \@minus.2\Cdp}%
  {.1\Cvs \@plus.3\Cdp}%
  {\reset@font\sffamily\normalsize}
}
\makeatother
\DeclareMathOperator*{\argmin}{argmin}

\usepackage{authblk}
\title{Multiple scattering ambisonics: three-dimensional sound field estimation using interacting spheres}
\author[1]{Shoken Kaneko}
\author[1]{Ramani Duraiswami}
\affil[1]{Perceptual Interfaces and Reality Laboratory, Computer Science and UMIACS, University of Maryland, College Park, MD 20742, USA}
\date{}                     
\DeclarePairedDelimiter\floor{\lfloor}{\rfloor}
\def \LambdaE{\Lambda^{\circ}}
\def \LambdaB{\Lambda^{\bullet}}

\def \LambdaP{\Lambda^{\mathrm{(P)}} } 

\def \pin{p_{\mathrm{in}}}
\def \ptot{p_{\mathrm{tot}}}
\def \Anmc{\mathscr{A}_{n}^{m}}

\def \AmatS{\mathbf{A}^{\bullet}}
\def \kv{\mathbf{k}}
\def \rv{\mathbf{r}}
\def \rvs{\mathbf{r}_\mathrm{s}}

\begin{document}

\maketitle
\begin{abstract}
    Rigid spherical microphone arrays (RSMAs) have been widely used in \emph{ambisonics}~\cite{gerzon1973periphony} sound field recording.
    While it is desired to combine the information captured by a grid of densely arranged RSMAs for expanding the area of accurate reconstruction, or \emph{sweet-spots}, this is not trivial due to inter-array interference.
    Here we propose \emph{multiple scattering ambisonics}, a method for three-dimensional ambisonics sound field recording using multiple acoustically interacting RSMAs.
    Numerical experiments demonstrate the sweet-spot expansion realized by the proposed method. 
    The proposed method can be used with existing RSMAs as building blocks and opens possibilities including higher degrees-of-freedom spatial audio.
\end{abstract}

\section{Introduction}

\begin{figure}
\centering
\includegraphics[width=12cm]{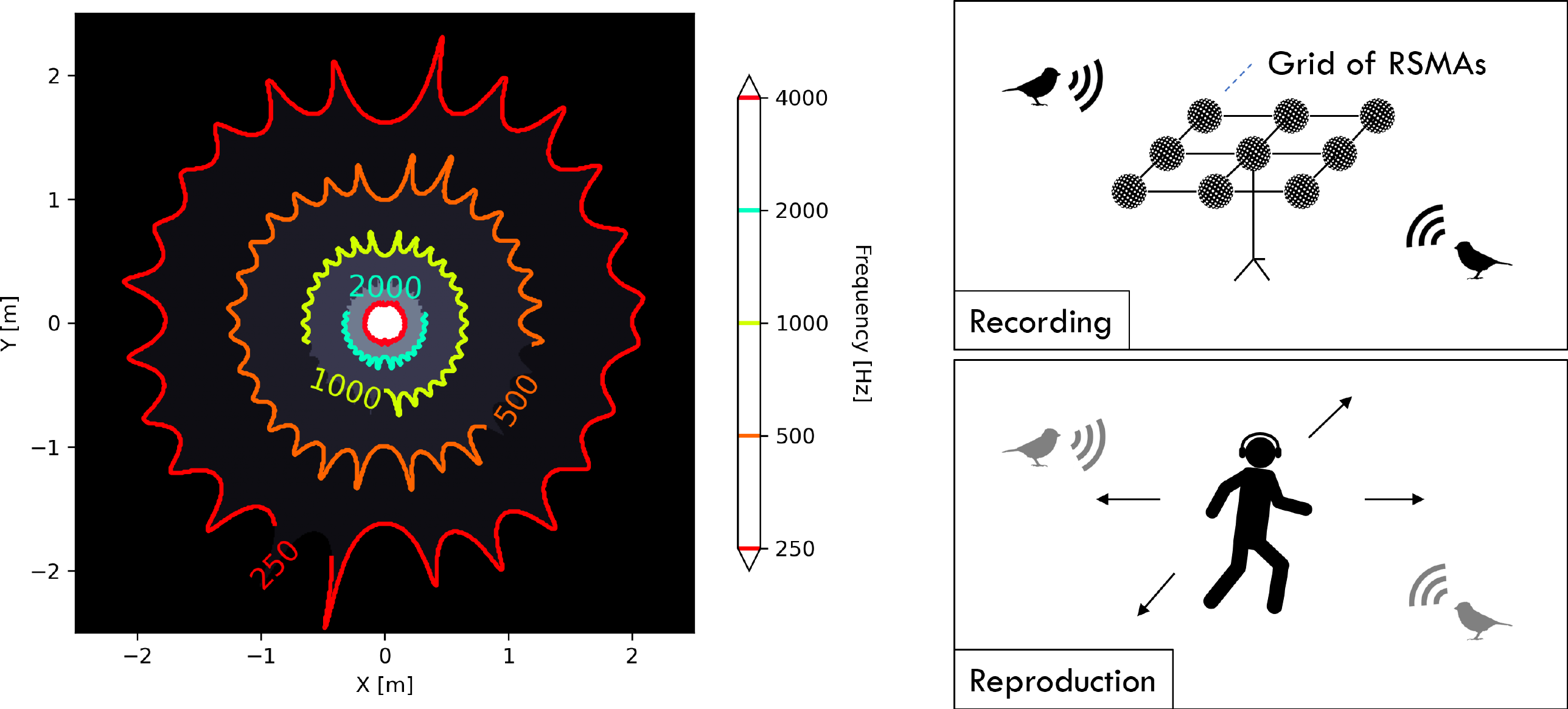}
\caption{Left: examples of reconstruction sweet-spots using a 252-channel RSMA for estimating incident fields with various frequencies. Details of the RSMA are described in \cref{section:experimentalResults}. The truncation degree of HOA which provides the largest sweet-spot is chosen for each frequency independently. Right: illustration of a sound field recording setup using MS-HOA (top) and its reproduction setup over headphones allowing translation of the listener (bottom).}
\label{fig:sss}
\end{figure}

Audio is indispensable in immersive technologies such as mixed reality (MR) and virtual reality (VR), which are receiving much attention.
For these applications, it is essential to develop technologies to capture, process, and render spatial sound fields with high precision for the presentation of truly realistic and immersive MR/VR experiences.
Ambisonics~\cite{gerzon1973periphony} as well as \textbf{higher-order ambisonics} (HOA)~\cite{daniel2003further}, which are established spatial audio  frameworks to capture, process and reproduce spatial sound fields based on its representation in the spherical harmonics domain, are receiving much attention due to the popularization of MR/VR platforms~\cite{youtube, facebook}, and its high compatibility with first-person view MR/VR.
Ambisonics spatial audio capturing and processing consists of a microphone array and signal processing that is used to encode the raw microphone array signal to the spherical harmonics-domain spatial description format, which is referred to the ambisonics signal.
This ambisonics signal is decoded to the signal which is fed to loudspeaker arrays to render the spatial sound field.
Such loudspeaker arrays are often virtualized by means of binaural technologies~\cite{noisternig20033d, zotkin2004rendering, kaneko2016ear} and played back via headphones.
Hence the high compatibility of ambisonics with MR/VR applications that usually use headphones for audio playback.
Due to its formulation in the spherical harmonics-domain, a typical implementation of an ambisonics recording device is employing a \textbf{spherical microphone array} (SMA)~\cite{gerzon1973periphony,meyer2002highly,daniel2003further,sakamoto2015sound,kaneko2018development}.
Often, SMAs are mounted on sound-hard spherical scatterers in order to avoid the instability arising in encoding filters for hollow microphone arrays due to singularities originating from the roots of the spherical Bessel function~\cite{daniel2003further}, and for its mechanical stability as hardware.
This form of a SMA is referred to as a \textbf{rigid SMA} (RSMA).
Despite its success in first-person immersive audio with only three \textbf{degrees-of-freedom} (DoF) which are associated with the rotation of the listener, ambisonics suffers from the diminishing size of the accurate reconstruction area, referred to as the \emph{sweet-spot}, as the frequency increases, hence limiting its efficacy in higher DoF spatial audio reproduction allowing translation of the listener.
This is visualized in \cref{fig:sss} (left), showing the resulting reconstruction sweet-spots for incident plane waves with various frequencies.
Here, the sweet-spot is defined as the region where the signal-to-distortion ratio (SDR) of the estimated field with respect to the ground truth incident field is above 30~dB.
In order to expand the sweet-spot of ambisonics reproduction, the simplest way is to develop RSMAs with larger number of microphones. 
Although this is an effective approach, it comes with a significant development and device cost. 
An alternative approach is to combine multiple existing RSMAs and integrate the captured information.
However, this is not a trivial task due to the inter-array interference. 
Here, \emph{multiple scattering higher-order ambisonics} (MS-HOA), a three-dimensional (3D) sound field capturing scheme using multiple RSMAs with fully considering inter-array interaction due to multiple scattering~\cite{martin2006multiple} is proposed.
Numerical experiments show that MS-HOA successfully creates sound field representations with expanded sweet-spots even when the RSMAs are densely arranged with small spacing, which is not achieved without the consideration of inter-array interaction.
An example sound field recording and reproduction setup allowing translation of the listener is illustrated in \cref{fig:sss} (right).

\section{Conventional ambisonics encoding using a single RSMA}
The conventional framework of ambisonics encoding using a single RSMA is briefly reviewed.
Ambisonics encoding and decoding can be performed by either relying on solving a linear system using least squares~\cite{daniel2003further} or relying on spherical harmonic transformation using numerical integration~\cite{poletti2005three}.
Since the first approach allows more flexibility of the microphone array configuration, this approach is adopted here.
In the present work, all formulations are presented in the frequency-domain, which can be converted to time-domain representations by inverse Fourier transform. All individual microphone capsules are assumed to be omnidirectional.
The spherical harmonics used are defined as 
\begin{equation}
Y _ { n } ^ { m } ( \theta , \varphi ) \equiv \sqrt { \frac { ( 2 n + 1 ) } { 4 \pi } \frac { ( n - m ) ! } { ( n + m ) ! } } P _ { n } ^ { m } ( \cos \theta ) e ^ { i m \varphi },
\end{equation}
with $\theta$ and $\varphi$ the polar and azimuthal angle, respectively, 
and $P_n^m(x)$ and $P_n(x)$ respectively the associated and regular Legendre polynomials:
\begin{equation}
P _ { n } ^ { m } ( x ) \equiv ( - 1 ) ^ { m } \left( 1 - x ^ { 2 } \right) ^ { m / 2 } \frac { d ^ { m } } { d x ^ { m } } \left( P _ { n } ( x ) \right), \quad P _ { n } ( x ) \equiv \frac { 1 } { 2 ^ { n } n ! } \frac { d ^ { n } } { d x ^ { n } } \left( x ^ { 2 } - 1 \right) ^ { n }.
\end{equation}
The above definition of spherical harmonics provides an orthonormal basis:
\begin{equation}
\int _ { \theta = 0 } ^ { \pi } \int _ { \varphi = 0 } ^ { 2 \pi } Y _ { n } ^ { m }( \theta , \varphi ) Y _ { n ^ { \prime } } ^ { m ^ { \prime }  }( \theta , \varphi )^{*} d \Omega = \delta _ { n n ^ { \prime } } \delta _ { m m ^ { \prime } },
\end{equation}
with $\delta _ { ij }$ the Kronecker delta.

\def \LambdaE{\Lambda^{\circ}}
\def \LambdaB{\Lambda^{\bullet}}
\def \LambdaP{\Lambda^{\mathrm{(P)}} } 
\def \pin{p_{\mathrm{in}}}
\def \pscat{p_{\mathrm{scat}}}
\def \ptot{p_{\mathrm{tot}}}
\def \ptotvec{\mathbf{p}_{\mathrm{tot}}}
\def \Snm{S_{n}^{m}}
\def \Rnm{R_{n}^{m}}
\def \Snms{S_{n}^{m(s)}}
\def \Snmst{S_{n}^{m(t)}}
\def \Rnms{R_{n}^{m(s)}}
\def \Anmc{A_{n}^{m}}
\def \Anmcest{A_{n}^{m(\mathrm{est})}}
\def \Anms{A_{n}^{m(s)}}
\def \Bnms{B_{n}^{m(s)}}
\def \Bnmt{B_{n}^{m(t)}}
\def \Ac{A}
\def \Avecest{\mathbf{A}^{\mathrm{(est)}}}
\def \Avec{\mathbf{A}}
\def \Avecin{{\mathbf{A}^{\mathrm{(in)}}}}
\def \Bvec{\mathbf{B}}
\def \AmatS{\mathbf{A}^{\bullet}}
\def \AmatSest{\mathbf{A}^{\bullet}_{\mathrm{(est)}}}
\def \Tf{\mathcal{T}_{\mathrm{F}}}
\def \Tfmat{T_{\mathrm{F}}}
\def \Timat{T_{\mathrm{I}}}
\def \kv{\mathbf{k}}
\def \rv{\mathbf{r}}
\def \rvs{\mathbf{r}_\mathrm{s}}
\def \Nc{N_{\mathrm{c}}}
\def \Ns{N_{\mathrm{S}}}
\def \Nq{N_{\mathrm{Q}}}
\def \Ncin{N_{\mathrm{c}}^{\mathrm{(in)}}}
\def \Ncfwd{N_{\mathrm{c}}^{\mathrm{(fwd)}}}
\def \Ncest{N_{\mathrm{c}}^{\mathrm{(est)}}}
\def \Lin{L^{\mathrm{(in)}}}
\def \Lfwd{L^{\mathrm{(fwd)}}}
\def \Lest{L^{\mathrm{(est)}}}
\def \qs{q^{(s)}}
\def \qvs{\mathbf{r}_{q}^{(s)}}
\def \rvt{\mathbf{r}_t}
\def \rvss{\mathbf{r}_s}
\def \Single{\textbf{Single}}
\def \HOA{\textbf{HOA}}
\def \MSHOA{\textbf{MS-HOA}}
\def \Amatin{\mathbf{A}^{\mathrm{(in)}}}
\def \Amatest{\mathbf{A}^{\mathrm{(est)}}}
\def \Linear{\textbf{Linear6}}
\def \Circular{\textbf{Circular9}}
\def \Cart{\textbf{Cartesian9}}

The process of obtaining the ambisonics signal $\Anmc(k)$, the weights of the spherical basis functions of the three dimensional sound field representing an arbitrary incident field of wavenumber $k$, from the signal captured by the microphone array is referred to as ambisonics \emph{encoding}.
An arbitrary incident field can be expanded in terms of the regular spherical basis functions $j_{n}(kr) Y_{n}^{m}(\theta, \varphi)$ of the three-dimensional Helmholtz equation in the spherical coordinate system $(r,\theta,\varphi)$:
\begin{equation}\begin{aligned}
\pin &= \sum_{n=0}^{\infty} \sum_{m=-n}^{n} \Anmc(k) j_{n}(kr) Y_{n}^{m}(\theta, \varphi),
\end{aligned}\label{eq:reconstruction}\end{equation}
with $j_{n}(x)$ the spherical Bessel function of degree $n$.
The total field $\ptot$, which is the sum of the incident field and the field scattered by a rigid sphere with radius $R$ located at $O$, the origin, is given by:
\begin{equation}\begin{aligned}
\ptot &= \sum_{n=0}^{\infty} \sum_{m=-n}^{n} \Anmc(k) \left\{ j_n(kr) - h_n(kr) \frac{j_{n}^{\prime}(kR)}{h_{n}^{\prime}(kR)} \right\}  Y_{n}^{m}(\theta, \varphi),\\
\end{aligned}\end{equation}
with $h_n(x)$ the spherical Hankel function of the first kind with degree $n$~\cite{morse1986theoretical}.
On the surface of the rigid sphere, i.e. $r=R$, this total field is evaluated as:
\begin{equation}\begin{aligned}
\ptot|_{r=R} &= \sum_{n=0}^{\infty} \sum_{m=-n}^{n} \Anmc(k)\frac{ i}{(kR)^{2} h_{n}^{\prime}(kR)} Y_{n}^{m}(\theta, \varphi).\\
\end{aligned}\end{equation}
The total field captured by the $q$-th microphone on the surface of the RSMA located at $(R,\theta_q,\varphi_q)$ is therefore given by:
\begin{equation}\begin{aligned}
\ptot^{(q)} &= \sum_{n=0}^{\infty }\sum_{m=-n}^{n}  \frac{ i }{(kR)^{2} h_{n}^{\prime}(kR)}  Y_{n}^{m}(\theta_q, \varphi_q) \Anmc(k).\\
\end{aligned}\end{equation}
By truncating the infinite series with $n= \Nc$, this result can be represented in the following vector form:
\begin{equation}\begin{aligned}
\mathbf{p}_{\mathrm{tot}} &= \Lambda \mathbf{A},\\
\end{aligned}\end{equation}
where $\mathbf{p}_{\mathrm{tot}}$ is a vector holding $\ptot^{(q)}$ in its $q$-th entry,
$\mathbf{A}$ is a vector holding $\Anmc(k)$ in its $(n^2+n+m+1)$-th entry,
and $\Lambda$ is the matrix holding $\frac{ i }{(kR)^{2} h_{n}^{\prime}(kR)}  Y_{n}^{m}(\theta_q, \varphi_q)$ in its $(q,n^2+n+m+1)$ entry.
The goal of ambisonics encoding is to obtain $\Anmc(k)$ for all $n$s and $m$s up to the truncation degree $n = \Ncin$, i.e. $0\le n \le \Nc$ and $|m|\le n$, from the observation $\mathbf{p}_{\mathrm{tot}}$.
This problem can be solved by regularized least squares with a minimization objective:
\begin{equation}\begin{aligned}
L_{\mathrm{enc}} &= ||\mathbf{p}_{\mathrm{tot}}-\Lambda \mathbf{A}||_2^2 + \sigma ||\mathbf{A}||_2^2,\\
\end{aligned}\end{equation}
with $\sigma$ a regularization parameter, and the solution given by:
\begin{equation}\begin{aligned}
\Avecest &= \argmin_{\mathbf{A}} L_{\mathrm{enc}} = (\Lambda^H\Lambda + \sigma I)^{-1}\Lambda^H \mathbf{p}_{\mathrm{tot}} = E \mathbf{p}_{\mathrm{tot}},
\end{aligned}\label{eq:sphericalHOAEncoding}\end{equation}
where $E \equiv (\Lambda^H\Lambda + \sigma I)^{-1}\Lambda^H$ is the regularized encoding matrix.

\section{Proposed method}

\def \rl{r_{\mathrm{long}}}
\def \rs{r_{\mathrm{short}}}

In the proposed method, a grid of multiple RSMAs is used to estimate $\pin$ \cref{eq:reconstruction}.
The goal of ambisonics encoding in MS-HOA is to estimate $\Anmc(k)$ \cref{eq:reconstruction}, 
for $0\le n \le \Nc$ and $|m|\le n$ from observations of the sound pressure at discrete microphone capsule positions mounted on the surfaces of multiple RSMAs. 
In the following, a system of $\Ns\ge2$ RSMAs where each RSMA has a radius $a_s$ is considered. 
Here, $s$ is the index of the RSMA. Hereafter, the argument $k$ is omitted from $\Anmc(k)$.

\subsection{The forward problem: multiple scattering due to an arbitrary incident field}
It is known that the problem of multiple scattering in a system of multiple spherical scatterers, i.e. computing the scattered field $\pscat$ given $\pin$ and the configuration of the scattering spheres, can be solved analytically~\cite{gumerov2002computation, martin2006multiple}.
This problem is referred to as the \emph{forward problem}.
The procedure of solving the forward problem is briefly described here.
First, $\Anmc$, the expansion coefficients at $O$ \cref{eq:reconstruction} truncated at degree $n=\Ncin$, are translated to the positions of the RSMAs using the translation operators $T^{(s,O)}_{R|R}$ resulting in the expansions
$\Avec^{(s)} = T^{(s,O)}_{R|R} \Avecin$,
where $\Avec^{(s)}$ and $\Avecin$ are vectors of length $\Lin \equiv (\Ncin+1)^2$ holding $\Anms$ and $\Anmc$ in its $(n^2+n+m+1)$-th entry, respectively.
The $\Avec^{(s)}$ coefficients are then further truncated at degree $n=\Ncfwd$.
Two distinct truncation numbers $\Ncin$ and $\Ncfwd$ are introduced here in order to achieve sufficient accuracy of the translation operation 
while limiting numerical error in the computation of the scattered field.
Given the set of expansions $\Anms$ at each RSMA position, the contribution of each RSMA to the scattered field $\Bnms$ can be computed by solving the linear system:
\begin{equation}\begin{aligned}
\Avec' = S \Bvec',
\end{aligned}\end{equation}
where $\Avec'$ and $\Bvec'$ are concatenations of $\Ns$ vectors $\{\Avec^{(1)}, \Avec^{(2)}, ..., \Avec^{(\Ns)}\}$ and $\{\Bvec^{(1)}, \Bvec^{(2)}, ..., \Bvec^{(\Ns)}\}$, where $\Bvec^{(s)}$ are vectors of length $\Lfwd \equiv (\Ncfwd+1)^2$ holding $\Bnms$ in its $(n^2+n+m+1)$-th entry.
$S$ is referred to as the \emph{system matrix}, which is a block matrix holding the inter-sphere translation operator $T^{(s,t)}_{S|R}$ from the $t$-th sphere to the $s$-th sphere in its off-diagonal $(s,t)$-block and the ``single scattering matrix" $\Lambda^{(s)}$ in its diagonal blocks:
\begin{equation}
S=
\left(\begin{array}{cccc}
\Lambda^{(1)}      & -T^{(1,2)}_{S|R}     & ...      & -T^{(1,\Ns)}_{S|R} \\
-T^{(2,1)}_{S|R}       & \Lambda^{(2)}      & ...          &  -T^{(2,\Ns)}_{S|R} \\
\vdots & \vdots & \ddots & \vdots\\
-T^{(\Ns,1)}_{S|R}      & -T^{(\Ns,2)}_{S|R}      & \dots  & \Lambda^{(\Ns)}\\
\end{array}\right),
\end{equation}
where $\Lambda^{(s)}$ is a diagonal matrix holding $-\frac{h'_n(ka_s)}{j'_n(ka_s)}$ in its $(l,l)$ entry with $l=n^2+n+m+1$.
The translation operators $T^{(s,O)}_{R|R}$ and $T^{(s,t)}_{S|R}$ can be computed by various methods, including explicit expressions based on Clebsch-Gordan coefficients or Wigner 3-j symbols~\cite{epton1995multipole}, or methods based on recurrence relations~\cite{chew1992recurrence}. 
The total field $\ptot$ evaluated at $\qvs$, the $q$-th microphone position belonging to the $s$-th RSMA, is the sum of the scattered field contributions from all the RSMAs and the incident field $\pin$:
\begin{equation}\begin{aligned}
\ptot(\qvs) &= \pscat(\qvs) + \pin(\qvs)\\
&= \sum_{n=0}^{\Ncfwd} \sum_{m=-n}^{n} \left( \sum_{t=1}^{\Ns}\Bnmt \Snmst(\qvs) + \Anms \Rnms(\qvs) \right),
\end{aligned}\end{equation}
where $\Rnms(\qvs)$ and $\Snmst(\qvs)$ are the regular and singular spherical basis functions expanded at the location of the $s$-th and $t$-th sphere, respectively, and evaluated at the position of the $q$-th microphone capsule belonging to the $s$-th RSMA:
\begin{equation}\begin{aligned}
\Rnms(\qvs) &= j_n(ka_s)Y_n^m(\qvs - \rvss), \quad \\
\Snmst(\qvs) &= h_n(k |\qvs - \rvt|) Y_n^m(\qvs - \rvt).
\end{aligned}\end{equation}
Alternatively, $\pin(\qvs)$ could be evaluated directly using $\Anmc$ instead of the translated $\Anms$ coefficients.
The whole procedure of the forward problem can be expressed by a linear operator $\Tfmat$ which is referred to as the forward operator:
\begin{equation}\begin{aligned}
\ptotvec &= \Tfmat \Amatin,\\
\end{aligned}\end{equation}
where $\mathbf{p}_{\mathrm{tot}}$ is a vector holding the values of $\ptot(\qvs)$.

\subsection{The inverse problem: MS-HOA encoding}
\begin{figure}
\centering
\includegraphics[width=12cm]{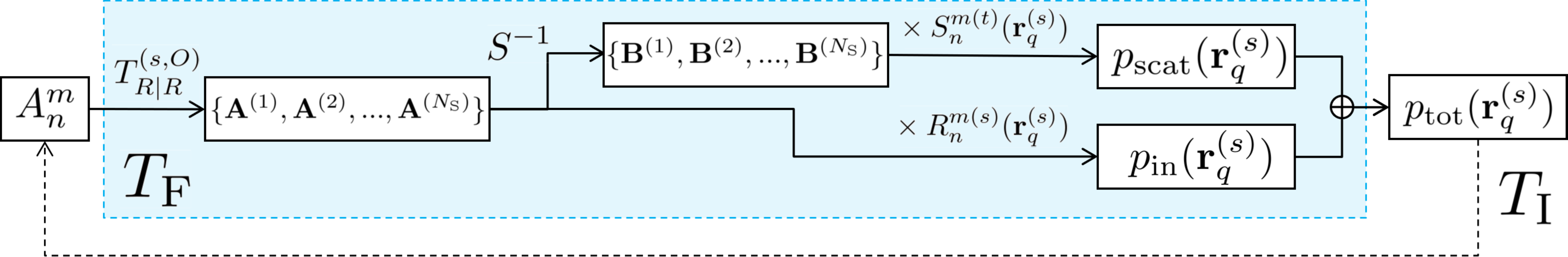}
\caption{The scheme of the forward problem, i.e. computation of the total sound field at the microphone positions given the incident field, and the inverse problem, i.e. estimation of the incident field given the sound pressure at the microphone positions.}
\label{fig:algo}
\end{figure}

The matrix representing $\Tfmat$ can be constructed by applying the operator to all bases up to $n\le \Ncin$.
The estimate of the incident field can then be obtained via regularized least squares:
\begin{equation}\begin{aligned}
\Amatest &= (\Tfmat^H\Tfmat + \sigma I)^{-1}\Tfmat^H \mathbf{p}_{\mathrm{tot}} = \Timat \mathbf{p}_{\mathrm{tot}},
\end{aligned}\label{eq:MSHOAEncoding}\end{equation}
where $\Amatest$ is a vector holding the estimated coefficients $\Anmcest$ in its $(n^2+n+m+1)$-th entry up to $n\le\Ncin$ 
and $\Timat \equiv (\Tfmat^H\Tfmat + \sigma I)^{-1}\Tfmat^H$ is the encoding matrix for MS-HOA with $\sigma$ a regularization parameter.
The scheme of the forward and inverse problem is summarized in \cref{fig:algo}.

\section{Numerical experiments}\label{section:experimentalResults}

\begin{figure}
\centering
\includegraphics[width=12cm]{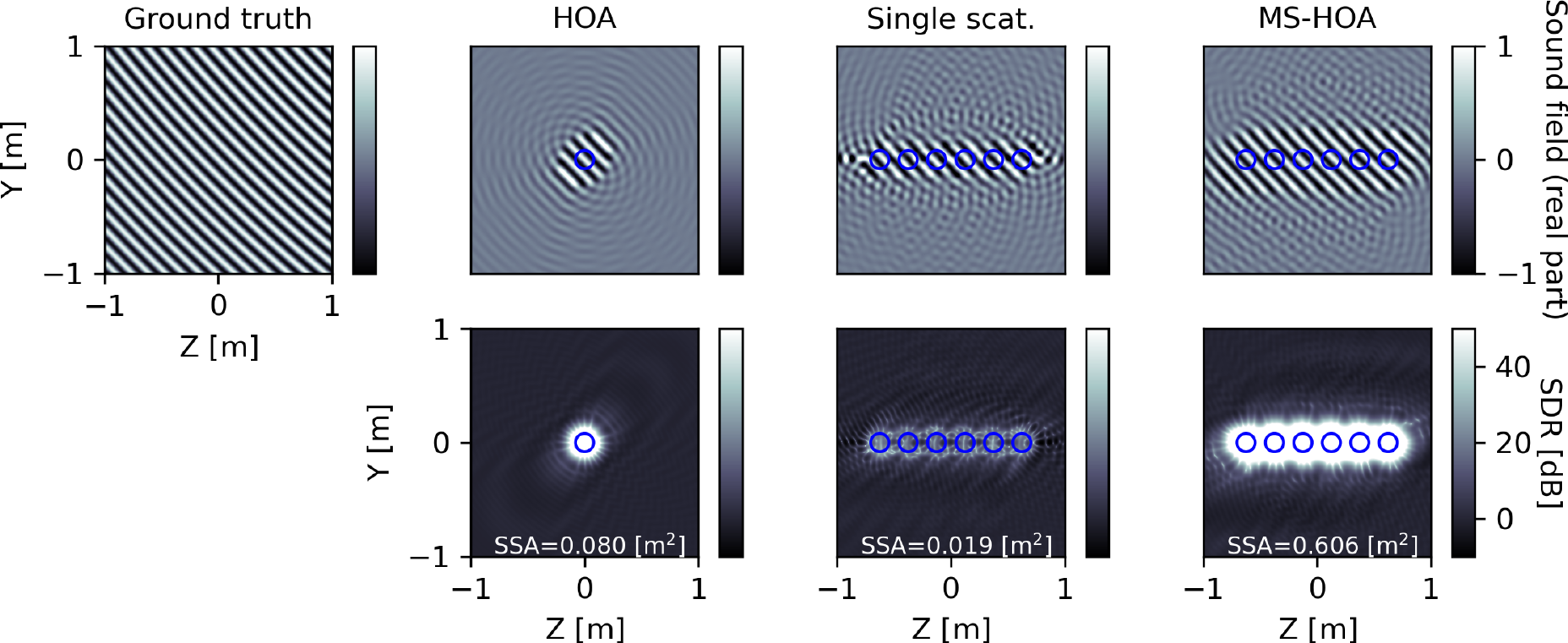}
\caption{Results for the linear 6-sphere RSMA grid for an incident field of 4~kHz. Top row from left to right: real part of the sound pressure for the ground truth incident field, incident field estimated by \HOA, \Single, and \MSHOA, respectively. Bottom row from left to right: SDR map of the estimated fields with respect to the ground truth field using \HOA, \Single, and \MSHOA, respectively. 
The blue circles represent the positions and sizes of the RSMAs.
The truncation number is $\Nc=14$ in \HOA, $\Ncin=55$ and $\Ncfwd=20$ in \Single~and \MSHOA.}
\label{fig:fields_lin6}
\end{figure}

\begin{figure}
\centering
\includegraphics[width=8cm]{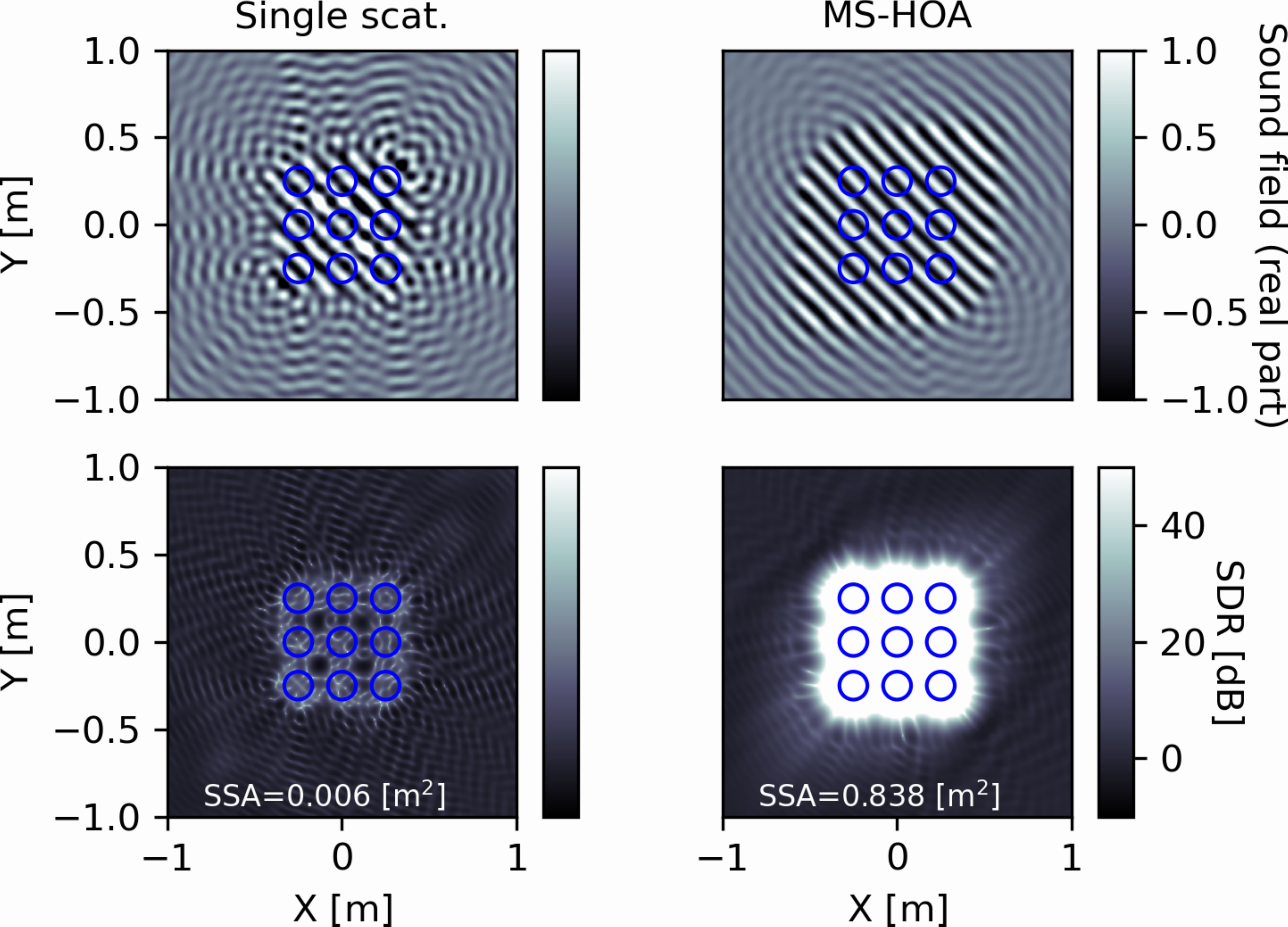}
\caption{Results for the regular Cartesian 9-sphere RSMA grid. The estimated incident field (top row) and SDR map (bottom row) for \Single~(left column) and \MSHOA~(right column), respectively.
The blue circles represent the positions and sizes of the RSMAs.
The same incident field as \cref{fig:fields_lin6} is used. The truncation numbers are $\Ncin=45$ and $\Ncfwd=16$.}
\label{fig:fields_cart9}
\end{figure}

MS-HOA recording and encoding into HOA coefficients was validated by numerical experiments.
Grids of RSMAs where each individual RSMA is a 252-channel SMA mounted on a rigid spherical scatterer with a radius of 8~cm are considered.
The spherical Fibonacci grid~\cite{swinbank2006fibonacci, kaneko2018development} of 252 points was used for the microphone capsule positions.
A real-world implementation of a 252-channel RSMA with a similar size has been demonstrated in the past~\cite{sakamoto2015sound}. 
As the RSMA grid, a linear grid of 6 RSMAs and a regular Cartesian grid of 9 RSMAs was used in the experiments. 
The spacing between the nearest neighbour RSMA was set to 25~cm.
The sound field generated by a monopole source located at $\rvs=(10\mathrm{m},10\mathrm{m},10\mathrm{m})$ was used as the incident field.
The signal captured by the grid of RSMAs was encoded into the HOA coefficients $\Amatest$ with the proposed method (\MSHOA).
While prior works on the forward problem report heuristics for choosing the parameter $\Ncfwd$, e.g. $\Ncfwd=\floor*{eka}$~\cite{gumerov2002computation}, here $\Ncfwd$ was treated as a free hyper-parameter.
The case where inter-sphere interaction is switched off, i.e. the method which only considers single scattering (\Single), 
and the case of conventional HOA encoding using only one building block RSMA (\HOA) 
are also computed as baselines.
The analytical reconstruction of the estimated incident field was computed by \cref{eq:reconstruction} and was compared to the ground truth incident field $\pin$ in terms of the SDR and the size of the reconstruction \textbf{sweet-spot area} (SSA) measured in the $xy$-plane or the $yz$-plane depending on the configuration of the RSMA grid.
The SSA is defined here as the total area where the SDR surpasses 30~dB, which is measured using the regular Cartesian grid points on a plane which correspond to the pixels in \cref{fig:fields_lin6}-\cref{fig:fields_cart9}.
The regularization hyperparameter was optimized by grid search independently for both the \Single~baseline and the proposed \MSHOA.
Regularization was not applied to the single sphere \HOA~baseline due to its minor effect for this case, while the truncation number $\Nc$ was chosen as the one providing the largest SSA for the given RSMA.
The results for the linear 6-sphere RSMA grid with a incident field frequency of 4kHz are shown in \cref{fig:fields_lin6}. 
The sweet-spot of reconstruction is successfully expanded with MS-HOA while the SSA and SDR is significantly degraded if only single scattering is considered.
The results for the regular Cartesian 9-sphere grid is shown in \cref{fig:fields_cart9}, demonstrating planar expansion of the sweet-spot.

\section{Related work and discussion}
Multiple scattering ambisonics, a method to capture 3D sound fields using multiple acoustically interacting RSMAs, was proposed.
MS-HOA allows to integrate the information captured by multiple densely arranged RSMAs and can be used to expand the reconstruction sweet-spots in 3D sound field reproduction.
The numerical experiments demonstrated that the proposed method successfully captures spatial sound fields with expanded reconstruction sweet-spots which was not possible without the consideration of inter-array interaction due to multiple scattering.

A related method using the translation of multipoles was introduced in~\cite{samarasinghe2014wavefield}.
This method was based on the assumption that the SMAs do not physically interact with each other, i.e. the SMAs do not cause scattering that affect other SMAs.
This assumption is violated if the SMAs are densely arranged RSMAs, which scatter the incident field and interact with each other by multiple scattering. 
As shown in the numerical experiments, the approach without considering inter-array interaction becomes inaccurate if the RSMAs are arranged with small spacing.
Recently, the consideration of inter-array multiple scattering has been demonstrated to improve the reconstruction accuracy in a two-dimensional sound field reconstruction problem using multiple cylindrical microphone arrays~\cite{nakanishi2019two}.
Two-dimensional modeling, however, is insufficient for modern spatial audio applications such as MR/VR where 3D audio representation and rendering is essential.
Our work enables the use of interacting rigid microphone arrays for 3D spatial audio.

The expanded reconstruction sweet-spots with linear or planar spreads realized by the proposed method could be useful in applications including sound field reproduction in theaters or in meeting rooms where the sweet-spot should cover multiple listeners sitting next to each other, or higher DoF MR/VR where the translation of the listener needs to be supported.
Developing techniques to reduce the cost of MS-HOA recording in terms of hardware, computation, and bandwidth is important for practical applications and are subjects of future research.

\section*{Acknowledgments}
Shoken Kaneko thanks to Japan Student Services Organization and Watanabe Foundation for support via scholarships.

\bibliographystyle{unsrt}
\bibliography{bib}

\end{document}